\documentclass[aps,pre,preprint,groupedaddress]{revtex4-1}
\usepackage{amssymb,amsfonts,amsmath}
\usepackage[dvips]{graphicx,color,psfrag}

\begin{document}
\title{Evolutionary game on networks with high clustering coefficient}

\author{Satoru Morita}
\email[]{morita@sys.eng.shizuoka.ac.jp}
%\homepage[]{Your web page}
%\thanks{}
\affiliation{Department of Mathematical and Systems Engineering, Shizuoka University, Hamamatsu, 432-8561, Japan}

\begin{abstract} 
This study investigates the influence of  lattice structure in
evolutionary games.
The snowdrift games is considered in networks 
with high clustering coefficients, that use four different 
strategy-updating.
Analytical conjectures using pair approximation were
compared with the numerical results.
Results indicate that general statements asserting that 
the lattice structure enhances cooperation are misleading. 
\end{abstract}

%\begin{keywords}
%Snowdrift game, evolutionary game theory, lattice structure, complex %networks, clustering coefficient
%\end{keywords}

\maketitle 

\section{Introduction}
Evolutionary games in complex networks have 
recently attracted attention in evolutionary biology,
behavioral science and statistical physics \cite{nowak0,lieberman,szabo}. 
One of the most important questions in these fields
is how network structure affects
the evolution of cooperative behavior \cite{nowak1,nowak2,ohtsuki,ohtsuki2,assenza,kuperman}.
Nowak and May noted that the lattice structure enhances
cooperative behavior  in the prisoner's dilemma game \cite{nowak1}.
Currently lattice structure is considered one of the mechanisms 
that support cooperation \cite{nowak0,nowakA}.
However, Hauert and Doebeli found
that lattice structure often inhibits cooperative behavior in
the snowdrift game \cite{hauert,doebeli}.
Thus, it is not clear how lattice structure
affects the evolution of cooperation in general situation.
Lattice structure are characteristically predisposed to high clustering. 
The purpose of this study was to establish a theoretical formula
that describes the influence of the clustering coefficient 
in evolutionary games.
Moreover, the effects of the  lattice structure. 
have been clarified 
The pair approximation technique was applied to obtain 
an analytical solution \cite{sato,morita}. 

The clustering coefficient is used to measure the tendency 
of nodes in a network to cluster together \cite{watts}.
The clustering coefficient of a single node
is defined as the probability that two 
randomly selected neighbors are connected to each other.
The clustering coefficient $C$ of the entire network
is determined by averaging the clustering coefficients of all nodes.
For many social networks, such as file actor collaborations \cite{watts}, telephone calls \cite{aiello},
e-mails \cite{ebel}, sexual relationships \cite{liljeros}, and
citation networks \cite{render}, the clustering coefficients
are greater than those of randomly established networks.
Although several studies have examined
the effects of clustering on the organization of cooperation \cite{assenza,kuperman},
there is little agreement as to
whether clustering 
promotes or inhibits the evolution of cooperation.
This study considers models with four different strategy-updating rules
and presents analytical predictions.

\section{Models}
Consider a static network with $n$ nodes.
An individual occupies each node.
Individuals play games with all neighbors and
their reproduction depends on the average payoff of a sequence of games.
The snowdrift game is considered as an example.
An individual chooses one of the
two strategies: cooperation (C) or defection (D).
The payoff matrix is given by \cite{hauert}
\begin{equation}
\begin{array}{cc}
&
\begin{array}{cc}
C \ \ \ & \ \ \ D
\end{array}\\
\begin{array}{c}
 C\\
 D
\end{array}
&
\left(
\begin{array}{cc}
 b -c/2& b-c \\
 b & 0
\end{array}
\right)
\end{array},
\label{eq_sdg}
\end{equation}
where the positive parameters $b$ and $c$ represent the benefit 
and cost of cooperation, respectively.
The cost-to-benefit ratio of mutual cooperation is defined by $r=c/(2b - c)$.
When $r<1$ (i.e., $b>c$), the snowdrift game has an inner Nash equilibrium, 
where the cooperator frequency is $1-r$.
In this case, the two strategies coexist in a well-mixed population.
This type of game is also known as the hawk-dove or chicken game.

Next, the networks
on which this evolutionary game is performed are defined.
All notes were assumed to have same degree $z$ (the number of
neighbors) to focus on the network cluster coefficient effects.
We used a random regular graph with a high clustering coefficient.
The edge exchange method \cite{kim} that selects
two links randomly and repetitively
was used to construct the graph.
The links were rewired only when the new network configuration 
was connected and had a larger clustering coefficient.
In addition,
three types of regular lattices with periodic boundary conditions were used:
square lattice with von Neumann
neighborhood ($z=4$), hexagonal lattice ($z=6$), and
square lattice with Moore neighborhood ($z=8$).
The clustering coefficient $C$ is calculated as zero for the von 
Neumann lattice,
although it is highly clustered.

The strategy was assumed to be updated 
stochastically and  asynchronously.
These are natural assumptions 
because strategy selection is
not  deterministic and occurs simultaneously in the population.
Four different strategy-updating rules were selected \cite{ohtsuki2,hauert}.
\begin{enumerate}
\item Birth-death (BD).
Choose an individual $i$ with probability 
proportional to its fitness $f_i$.
Then, choose another individual $j$ among the neighbors of 
individual $i$. 
Individual $j$ adopts the strategy of individual $i$.
\item Death-birth (DB).
Choose an individual $i$ at random.
Then, choose another individual $j$ among the neighbors of individual $i$
with probability proportional to fitness $f_j$.
Individual $i$ adopts the strategy of individual $j$.
\item Imitation (IM).
Choose an individual $i$ at random.
Then, choose another individual $j$ among individual $i$ and 
its neighbors with probability 
proportional to its fitness.
Individual $i$ adopts the strategy of individual $j$.
\item Local competition (LC). 
Choose an individual $i$ at random and 
then choose another individual $j$ among its neighbors randomly. 
Individual $j$ adopts the strategy of individual $i$
with probability $f_i/(f_i+f_j)$. 
\end{enumerate}
The fitness $f_i$ of individual $i$ is 
given by $1-w+w P_i$, where $P_i$ is 
the average payoff of all its neighbors.
The parameter $w$ is the intensity of selection \cite{nowak0,ohtsuki2}.
We assumed a weak selection with small $w$.
This weak selection assumption 
allowed the pair approximation to be performed analytically
within reason of what occurs in the biological world.

\section{Analytical Predictions}
A pair approximation was used to
calculate the equilibrium state.
Let $p_C$ and $p_D$ be the densities of cooperators (C) and
defectors (D), respectively.  
The pair densities taken into consideration $p_{CC}$, $p_{CD}$ and $p_{DD}$
represented the frequency that 
two neighboring pairs were CC, CD or DD.
Pairs CD and DC were not distinguished from each other.
Thus, $p_{CC}+p_{DD}+p_{CD}=1$.
The conditional probabilities $p_{C|C}$ and $p_{D|D}$ is given by
\begin{equation}
p_{C|C} =  p_{CC}/p_{C},  \
p_{D|D} =  p_{DD}/p_{D}.
\label{eq_p22}
\end{equation}
Considering 
$p_{C}=p_{{CC}}+p_{{CD}}/2$ and 
$p_{D}=p_{{DD}}+p_{{CD}}/2$,
the densities $p_C$, $p_D$, $p_{CC}$, $p_{CD}$ and $p_{DD}$
are represented as functions of $p_{C|C}$ and $p_{D|D}$:
\begin{eqnarray}
p_C & = &  \frac{1-p_{D|D}}{2-p_{C|C}-p_{D|D}} ,\nonumber\\
p_D & = &  \frac{1-p_{C|C}}{2-p_{C|C}-p_{D|D}} ,\nonumber \\
p_{CC} & = &  \frac{p_{C|C}(1-p_{D|D})}{2-p_{C|C}-p_{D|D}} ,\label{eqx} \\
p_{CD} & = &  \frac{2(1-p_{C|C})(1-p_{D|D})}{2-p_{C|C}-p_{D|D}} ,\nonumber \\
p_{DD} & = &  \frac{p_{D|D}(1-p_{C|C})}{2-p_{C|C}-p_{D|D}}\nonumber.
\end{eqnarray}
A triplet, which includes three nodes, can 
be one of two different configurations.
The triad has a node connected with two other nodes 
that do not connect with each other.
The triangle configuration is when all three nodes connected.
The standard pair approximation for a triad \cite{sato} leads to
\[
\begin{array}{lcl}
 p_{\angle CCD}&\approx&\displaystyle \frac{P_{CC}P_{CD}}{2P_C}\\
 p_{\angle CDD}&\approx&\displaystyle \frac{P_{DD}P_{CD}}{2P_D}.
\end{array}
\] 
However,
the use of an extended pair approximation \cite{morita} 
for a triangle provides
\[
 p_{{\triangle CCD}}:
 p_{{\triangle CDD}} \approx \frac{p_{{CC}}p_{{CD}}p_{{CD}}}
{p_{{C}} p_{{C}} p_{{D}}}:
 \frac{p_{{CD}} p_{{CD}} p_{{DD}}}{p_{{C}}p_{{D}}p_{{D}}}=  p_{C|C}:p_{D|D} .
\]
The first approximate equality was calculated from the 
Kirkwood superposition approximation \cite{kirkwood,matsuda2000}.
Therefore, we obtained 
\begin{equation}
\begin{array}{lcl}
 p_{C|CD}&\approx&\displaystyle (1-C)p_{C|C}+C\frac{p_{C|C}}{p_{C|C}+p_{D|D}},\\
 p_{D|DC}&\approx&\displaystyle (1-C)p_{D|D}+C\frac{p_{D|D}}{p_{C|C}+p_{D|D}},\\
\end{array}
\label{EPA}
\end{equation}
where $C$ represents the clustering coefficient,
$p_{C|CD}$ is the probability that
a neighbor of the end cooperator of a CD pair is
a cooperator, and $p_{D|DC}$ is the probability
that a neighbor of the end defector of a CD pair
is a defector.

A cooperator can become a defector only
when at least one defector exists in the neighborhood of the cooperator, 
and vice versa for all four strategy-updating rules
Thus, a strategy can be replaced only in CD pairs.
In the strategy-updating cases of BD and LC, 
the probability to choose C among a CD pair
is proportional to the average fitness
\begin{equation}
%P_{D\to C} \propto 
1-w+w[b-c+\frac{c}{2}(1-1/z)p_{C|CD}],
\label{EC1c}
\end{equation}
while
the probability to choose D is proportional to
\begin{equation}
%P_{C\to D} \propto 
1-w+w[b-b(1-1/z)p_{D|DC}].
\label{EC1d}
\end{equation}
The necessary condition for equilibrium is
that (\ref{EC1c}) equals to (\ref{EC1d}).
This condition is simplified as
\begin{equation}
\frac{c}{2}\ p_{C|CD}+b\ p_{D|DC}=\frac{c}{1-1/z}
\label{EC1}
\end{equation}
If eq.\  (\ref{EC1}) is correct,
the strategy changing rates coincide with each other: 
\begin{equation}
P_{C\to D}=P_{D\to C}.
\label{EC1all}
\end{equation}

In addition, in the equilibrium state,  
the rate at which CD pairs become CC needs to 
equal the rate at which CC pairs become CD.
The rate of change of the doublet density is given by
\begin{eqnarray}
P_{CD\to CC}&=&[1+(z-1)(1-p_{D|DC})]P_{D\to C}\\
\label{EC2a}
P_{CC\to CD}&=&(z-1)p_{C|CD}P_{C\to D}.
\label{EC2b}
\end{eqnarray}
Since $P_{CD\to CC}=P_{CC\to CD}$ and eq.\ (\ref{EC1all}),
another condition is
\begin{equation}
p_{C|CD}+p_{D|DC}=\frac{z}{z-1}.
\label{EC2}
\end{equation}
%It is independent of the updating rule.
Solving the system of eqs.\ (\ref{EC1}) and (\ref{EC2}),
yields the equilibrium solutions of $p_{C|CD}$ and $p_{D|DC}$.
Using (\ref{eqx}) and (\ref{EPA}), the cooperator equilibrium density
was calculated as
\begin{equation}
p_c=-\frac{z-C(z-1)}{z-2-C(z-1)}\left(r-\frac{1}{2}\right)+\frac{1}{2}.
\label{sol_bd}
\end{equation}
The result (\ref{sol_bd}) is valid for BD and LC.
\begin{figure}[t]
\begin{center}
\includegraphics[width=0.7\textwidth]{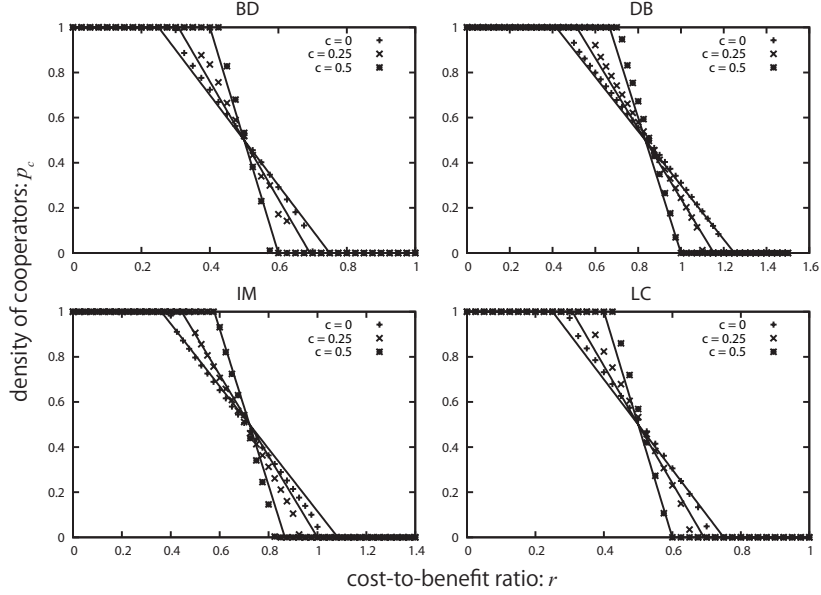}
\end{center}
\caption{The density $p_c$ of cooperators
plotted as a function of the cost-to-benefit ratio $r=c/(2b-c)$
for four different updating rules.
The clustering coefficient was set to $C=0$, $0.25$, and $0.5$
for fixed  $z=4$ and $w=0.5$.
The system size was 10,000.
In all simulations, $p_c$ was obtained by averaging 
the last 10,000 time steps after 
the first 10,000 ones, and each point resulted from 10 
different realizations.
The lines represent the predictions 
(\ref{sol_bd}) for BD and LC, 
(\ref{sol_db}) for DB, and (\ref{sol_im}) for IM. 
}\label{fig_1}
\end{figure}
 
It is more complicated to obtain a pair approximation 
that involves the effect of triangles in the case of DB and IM updating
The condition (\ref{EC2}) is also valid in these cases.
The following are conjecture equations
\begin{equation}
p_c=-\frac{[z-C(z -1)](z-1)}{[z-2 -C ( z-1)] (z+1)}\left(r-
\frac{1}{2}-\frac{1}{z-1}\right)+\frac{1}{2}
\label{sol_db}
\end{equation}
for DB updating, and 
\begin{equation}
p_c=
-\frac{[z-C(z -1)] (z-1)}{[z-2 -C (z-1)] (z+1)}\left[r-
\frac{1}{2}-\frac{z}{(z+2)(z-1)}\right]+\frac{1}{2}
\label{sol_im}
\end{equation}
for IM updating.
Although eqs.~(\ref{sol_db}) and (\ref{sol_im})  can be calculated by analogy 
with (\ref{sol_bd}). proper deviation do not exist yet.

\section{Numerical Results}
\begin{figure}
\begin{center}
\includegraphics[width=0.7\textwidth]{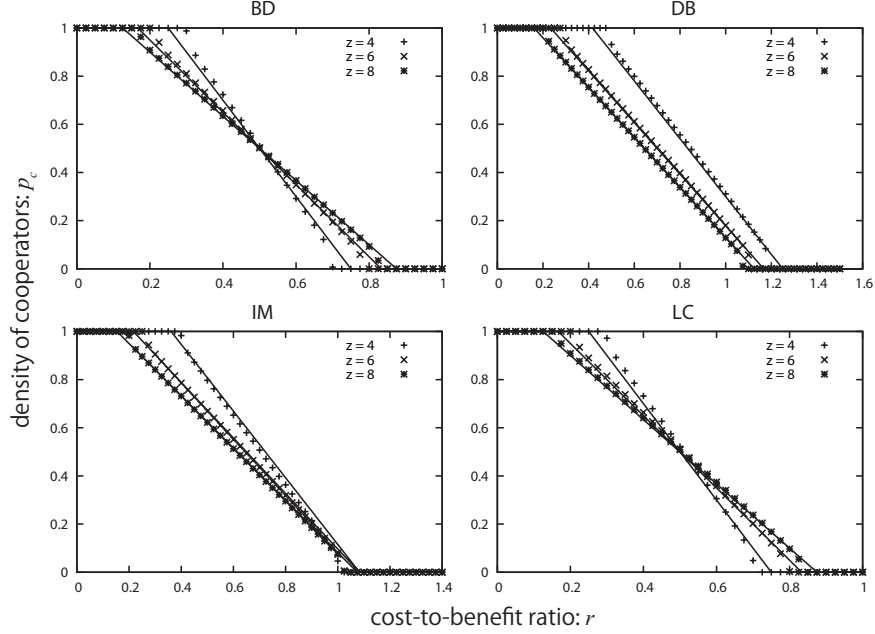}
\end{center}
\caption{
The density $p_c$ of cooperators
plotted for four different updating rules.
The degree was set to $z=4$, $6$ and $8$
for fixed  $C=0$ and $w=0.5$.
Lines represent predictions 
(\ref{sol_bd}) for BD and LC, 
(\ref{sol_db}) for DB, and (\ref{sol_im}) for IM. 
Other parameter values are the same as in Fig.~\ref{fig_1}
}\label{fig_2}
\end{figure}.
\begin{figure}
\begin{center}
\includegraphics[width=0.7\textwidth]{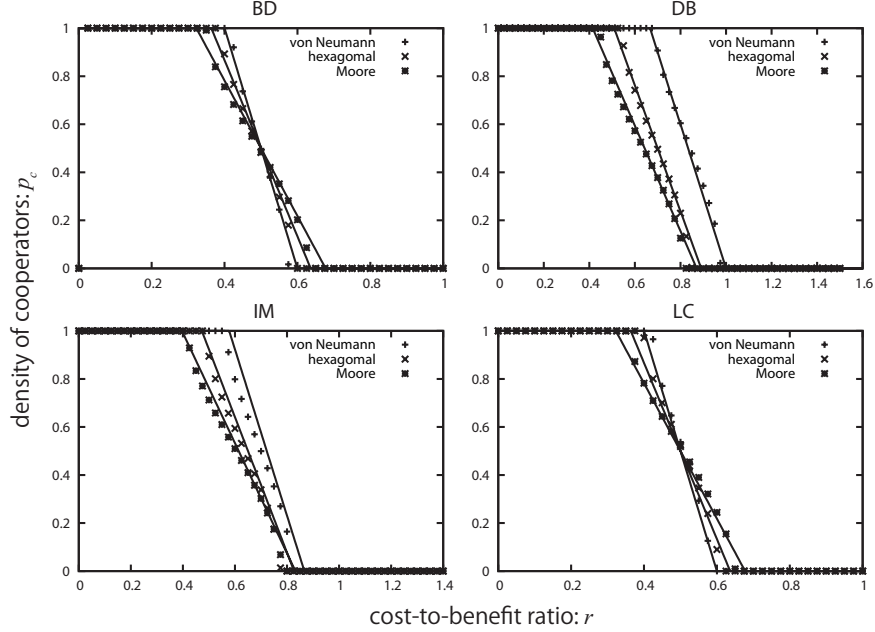}
\end{center}
\caption{The density $p_c$ of cooperators
plotted for four different updating rules.
The simulations were performed for the von Neumann square lattice ($z=4$),
hexagonal lattice ($z=6$) and Moore lattice ($z=8$).
Reference lines are superimposed 
(\ref{sol_bd}) for BD and LC, 
(\ref{sol_db}) for DB, and (\ref{sol_im}) for IM,
where the clustering coefficient $C$
was set to $C=0.5$ for the von Neumann square lattice,
$C=0.65$ for the hexagonal lattice, and 
$C=0.7$ for the Moore lattice.
Other parameter values are the same as in Fig.~\ref{fig_1}
}\label{fig_3}
\end{figure}

To confirm the predictions presented in the previous section, 
numerical results were performed for random networks with high clustering coefficients
(Figs.~\ref{fig_1} and \ref{fig_2}).
The predictions agree well with the numerical results.
Figure~\ref{fig_1}
shows that when the clustering coefficient increases,
the frequency of the major strategy increases for all four updating rules. 
Predictions (\ref{sol_bd}), (\ref{sol_db}), and (\ref{sol_im}) suggest
the interval of the coexisting region is
$[z-2-C(z-1)]/[z-C(z-1)]$ for BD and LC, 
and$\{[z-2-C(z-1)](z+1)\}/\{[z-C(z-1)](z-1)\}$ for DB and IM.
Thus, if $C>(z-2)/(z-1)$ the only one strategy can survive
for all four updating rules.
Figure~\ref{fig_2}
shows that 
the frequency of the majority increases when 
the degree $z$ decreases for BD and LC updating rules.
Cooperation is enhanced for small $z$ for DB and IM updating rules.

Figure \ref{fig_3} shows numerical results for 
three types of two-dimensional lattices.
The predictions
(\ref{sol_bd}) for BD and LC, 
(\ref{sol_db}) for DB, and (\ref{sol_im}) for IM,
were superimposed for reference
where the parameters were
set as $z=4$ and $C=0.5$ for the von Neumann lattice,
$z=6$ and $C=0.65$ for the hexagonal lattice, and 
$z=8$ and $C=0.7$ for the Moore lattice in Fig.\ref{fig_3}.
This result suggests 
that the `effective' cluster coefficients
are approximately 0.5, 0.65, and 0.7 
rather than the ``nominal'' values
0, 0.4, and 0.43.
The clustering coefficient measures the density of triangles
in a network. 
This deviation is because of 
the effect of loops of length four and above. 
Cooperator density increases when
the degree $z$ decreases for DB and IM updating rules.

\section{Discussion and Conclusion}
In conclusion, the frequency of the majority
increases with the clustering coefficient. 
In situations where cooperators and defectors coexist
and cooperators are the majority,
clustering enhances cooperative behavior.
When cooperators are the minority where cooperators and defectors coexist, clustering inhibits cooperative behavior.
These results are independent of the strategy-updating rule.
We can explain this tendency intuitively by using the heterophilicity \cite{park}
as follows.
From eqs.\ (\ref{eqx}), (\ref{EPA}) and (\ref{EC2}), 
the heterophilicity \cite{park} is calculated as
\begin{equation}
H:=\frac{p_{CD}}{2p_C p_D}=1-\frac{1}{(z-1)(1-C)},
\end{equation}
for all four updating rules.
It is obvious $H<1$, meaning that  
C have more connections to D than expected randomly.
When the clustering coefficient $C$ increases,
heterophilicity $H$ decreases.
In this case, the population is  more
exclusive, and it is more difficult for strategies to coexist. 
Consequently, the parameter region where two strategies can coexist
becomes narrow.
We performed numerical simulations for a
small world network \cite{watts} and a
scale-free network on geographical space \cite{morita2} (not shown)
to confirm the generality of this result, 
Essentially the same results were obtained.
Unfortunately, a rigorous derivation of
 (\ref{sol_db}) and (\ref{sol_im}) is not provided 
and it remains an open problem.

Lastly, we considered the prisoner's dilemma game,
where the payoff matrix is \cite{nowakA}
%\cite{christian}
\begin{equation}
\begin{array}{cc}
&
\begin{array}{cc}
C \ & \ \ D
\end{array}\\
\begin{array}{c}
C\\
D
\end{array}
&
\left(
\begin{array}{cc}
 b -c& -c \\
 b & 0
\end{array}
\right)
\end{array}.
\label{eq_pdg}
\end{equation}
In this case, mutual defection is the only strong Nash equilibrium,
regardless of the values of the parameters $b$ and $c$.
Thus, only defectors can survive in well-mixed population.
In addition, cooperators cannot survive for BD and LC updating rules.
The standard pair approximation for the DB updating rule
shows if $b/c>z$, only cooperators 
can survive; conversely, if $b/c<z$, only defectors
can survive \cite{ohtsuki}.
The result is the same in the case of IM updating,
except the threshold is $b/c=z+2$.
In any case, there is no parameter region where 
the two strategies can coexist.
Thus, the clustering coefficient  has no influence on
the density of cooperators in the prisoner's dilemma game.
In conclusion, the assertion that lattice structure enhances 
cooperative behavior is misleading.

%%\section*{Acknowledgments}
\acknowledgments
This work was supported by Grant-in-Aid for Scientific Research (No. 26400388) and CREST, JST.
Some of the numerical calculations were performed
on machines at the YITP of Kyoto University.

%\section*{Appendices}


\begin{thebibliography}{99}
\bibitem{nowak0}
M. A. Nowak,
\textit{Evolutionary Dynamics: Exploring the Equations of Life}
Harvard University Press, Cambridge, 2006.

\bibitem{lieberman}
E Lieberman, C Hauert, MA Nowak,
``Evolutionary dynamics on graphs,''
Nature {\bf 433}, pp.\ 312--316, 2005.

\bibitem{szabo}
Gy\"{o}rgy Szab\'{o}, G\'{a}bor F\'{a}th,
``Evolutionary games on graphs,''
Physics Reports {\bf 446}, pp.\ 97-216, 2007.

\bibitem{nowak1}
M.A. Nowak and R.M.  May, 
``Evolutionary games and spatial chaos,''
Nature {\bf 359}, 826--829, 1992.

\bibitem{nowak2}
M.A. Nowak and K. Sigmund,
``Evolutionary Dynamics of Biological Game,''
Science  {\bf 303}, 763--799 (2004).

\bibitem{ohtsuki}
H. Ohtsuki, C. Hauert, E. Lieberman, M. A. Nowak,
``A simple rule for the evolution of cooperation on graphs and social networks,''
Nature {\bf 441}, pp.~502--505, 2006.

\bibitem{ohtsuki2}
Hisashi Ohtsuki, Martin A Nowak,
``Evolutionary games on cycles,''
Proc. R. Soc. Lond., B, Biol. Sci, 
{\bf 273} pp.~2249--2256, 2006.

\bibitem{assenza}
S. Assenza, J. G\'{o}mez-Garde\~{n}es, and V. Latora,
``Enhancement of cooperation in highly clustered scale-free networks''
Phys. Rev. E {\bf 78}, 017101, 2008

\bibitem{kuperman}
M. N. Kuperman and S. Risau-Gusman,
``Relationship between clustering coefficient and the success of cooperation in networks''
Phys. Rev. E {\bf 86}, 016104, 2012

\bibitem{nowakA}
M. A. Nowak,
``Five rules for the evolution of cooperation''
Science {\bf 314}, pp.~1560--1563, 2006.

\bibitem{hauert}
Ch. Hauert and M. Doebeli,
``Spatial structure often inhibits the evolution of cooperation in the
snowdrift game,''
Nature {\bf 428}, pp.~643--646, 2004.

\bibitem{doebeli}
M. Doebeli and Ch. Hauert, 
``Models of cooperation based on the Prisoner's Dilemma 
and the Snowdrift game,''
Ecol. Lett. {\bf 8}, pp.~748--766, 2005.

\bibitem{sato}
K. Sato, H. Matsuda and A. Sasaki,
``Statistical Mechanics of Population,''
Prog. Theor. Phys. {\bf 88}, pp.~1035--1049, 1992.

\bibitem{morita}
S. Morita,
``Extended Pair Approximation of Evolutionary Game
on Complex Networks'' 
Prog. Theor. Phys. {\bf 119}, pp.~29-38, 2008.

\bibitem{watts}
D. J. Watts and S. Strogatz,
``Collective dynamics of 'small-world' networks,''
Nature {\bf 393}, pp.~440--442, 1998.

\bibitem{aiello}
W. Aiello, F. Chung, and L. Lu,
``A random graph model for massive graphs in Proceedings,''
the 32nd Annual ACM Symposium on Theory of Computing, 
pp.\ 171--180, 
Association of Computing Machinery, New York, 
2000.

\bibitem{ebel}
K. Ebel, L. -I. Mielsch and S. Bornholdt, 
``Scale-free topology of e-mail networks,''
Phys. Rev. E {\bf 66}, 035103, 2002.

\bibitem{liljeros}
F. Liljeros, C. R. Edling, L. A. N. Amaral, H. E. Stanley and Y. Aberg, 
``The web of human sexual contacts,''
Nature {\bf 411}, pp.\ 907--908, 2001.

\bibitem{render}
S. Redner,
``How popular is your paper? An empirical study of the citation distribution,''
Eur. Phys. J. B {\bf 4}, pp.\ 131--134, 1998.

\bibitem{christian}
H. Christian, M. A. Nowak, and K. Sigmund,
``Evolution of extortion in Iterated Prisoner's Dilemma games,''
Proc. Natl. Acad. Sci. USA {\bf 110}, pp.\ 6913--6918, 2013.

\bibitem{kim}
B. J. Kim, 
`` Performance of networks of artificial neurons: The role of clustering,''
Phys. Rev. E {\bf 69}, 045101(R), 2004.

\bibitem{kirkwood}
J. G. Kirkwood,
``statistical mechanics of fluid mixtures,''
J. of Chem. Phys. {\bf 3}, pp.\ 300--313, 1935.

\bibitem{matsuda2000}
H. Matsuda,
``physical nature of higher-order mutual information:
Intrinsic correlations and frustration,''
Phys. Rev. E {\bf 62}, pp.\ 3096--3102, 2000.

\bibitem{park}
J. Park and A. -L. Barab\'{a}si,
``Distribution of node characteristics in complex networks,''
Proc. Natl. Acad. Sci. USA {\bf 104}, pp.\ 17916--17920, 2007.

\bibitem{morita2}
S. Morita,
``Crossovers in scale-free networks on geographical space,''
Phys. Rev. E {\bf 73}, 035104, 2006. 
\end{thebibliography}
\end{document}